\documentstyle[12pt,psfig]{article}
\def\reference{\parskip 0pt\par\noindent\hangindent 0.5 truecm}
\def\kms{km ${\rm s}^{-1}$}
\def\HI{H\,{\sc i}}
\def\HII{H\,{\sc ii}}
\def\Msol{$M_\odot$}
\begin{document}
\title{\HI\  in  Early-Type Galaxies}

\author{Tom Oosterloo$^{1}$ \and
 Raffaella Morganti$^{2}$ \and
 Elaine Sadler$^{3}$
} 

\date{}
\maketitle

{\center
$^1$ Istituto di Fisica Cosmica, CNR, Via Bassini 15, 20133 Milan, Italy\\
toosterl@ifctr.mi.cnr.it\\[3mm]
$^2$ Istituto di Radioastronomia, CNR, Via Gobetti 101, 40129 Bologna, Italy\\
rmorgant@ira.bo.cnr.it\\[3mm]
$^3$ School of Physics,  University of Sydney, Australia, NSW 2006  \\
ems@physics.usyd.edu.au \\[3mm]
}

\begin{abstract}

\noindent 
We summarize the \HI\   properties of early-type  galaxies, in  particular the
differences in \HI\ morphologies  observed in early-type galaxies of different
luminosities. We  find that in  low-luminosity early-type galaxies the \HI\ is
almost always in a disk-like   structure, with central surface densities  high
enough for star formation to occur. In a few  luminous early-type galaxies the
\HI\ is also  in a disk or  in  a ring-like  structure,  but in  most luminous
early-type   galaxies the \HI\   has  a  relatively irregular  morphology. The
surface densities in the \HI\ disks in luminous  early-type galaxies are lower
than  in   the  \HI\ disks in    low-luminosity  early-type galaxies    and no
large-scale  star formation  should occur  in  these  disks.  We discuss these
different  \HI\   characteristics  in  the  context   of other  properties  of
early-type galaxies that correlate with luminosity.
\end{abstract}


\bigskip

\section{Introduction}

Early-type galaxies constitute quite a heterogeneous  group of galaxies.  Many
of  the  properties of   these systems vary  from  galaxy  to galaxy and, more
interestingly, vary systematically  with luminosity and environment.  Many  of
the differences may be related to the  amount of gas (and dissipation) present
during the formation and the evolution of the  galaxies. Moreover, evidence is
accumulating now that some early-type    galaxies probably have a   long-lived
interstellar medium  (ISM;   e.g.\  Knapp  1998).  To  help   understand   the
mechanisms and processes  behind the differences  between different early-type
galaxies and to study  the ISM observed in some  of these galaxies, it may  be
worthwhile to study the systematics of the  properties of the neutral hydrogen
in early-type galaxies as function of luminosity and environment. To this end,
we have  observed  a large number   of early-type galaxies  with the Australia
Telescope Compact Array and the  Very Large Array  (Morganti et al.\ 1997a,b,
1998a,b).  In particular, we have  tried to observe  galaxies over a range of
luminosities in  order to  see whether the  \HI\  properties of low-luminosity
early-type   galaxies (which we  define here  as   galaxies with absolute blue
magnitude in the range of $-16$ to $-19$) vary systematically with luminosity.
Here we give a brief overview of the results obtained so far.  In section 2 we
give  a brief overview of properties  of early-type galaxies observed at other
wavebands that   are relevant for  the discussion.   In  section  3 we briefly
discuss the \HI\ content of early-type galaxies and in  section 4 we summarize
the morphology  and the kinematics  of the \HI\ as  function of luminosity and
environment and  the relation of these  properties to other properties  of the
galaxies.

\begin{table}
\begin{center}
\begin{tabular}{lccc} 
\hline
\hline
Galaxy Type & Observed & Detected & \% \\
\hline
 E               &    64 & 3   &  5  \\
 E/S0            &    23 & 4   & 17  \\
 S0, SB0         &   103 & 21  & 20  \\
 Pec E and S0    &    20 & 9   & 45  \\
 S0/a and pec    &    35 & 15  & 43  \\
 Sa and pec      &   103 & 78  & 76  \\ 
\hline
\hline
\end{tabular}
\end{center}
\caption{\HI\ detection rates for early-type galaxies (from Bregman et al.\
1992)}
\end{table}

\section{Properties at other wavelengths}

There  are several properties of  early-type galaxies that systematically vary
with luminosity:

{\sl Stellar rotation}.   Some early-type galaxies  are `pressure  supported',
while in other galaxies the rotation of the stellar component is important for
the dynamics of the  system.   In general,   in luminous galaxies  the  random
motions dominate  while in lower    luminosity  systems the stellar   rotation
becomes more important (e.g., Davies et al.\ 1982).

{\sl Isophotal  shape}.  In  many early-type  galaxies, the isophotes  are not
perfect ellipses. In lower luminosity  galaxies, the isophotes are more  often
disky, while in luminous systems they tend to be more often box shaped.

{\sl Core properties}.  Imaging studies  performed  with {\sl  HST} have shown
that the central  density  distributions vary systematically  with  luminosity
(e.g., Lauer 1997).   Low-luminosity systems usually  have  steeper cores that
more luminous systems

{\sl  Excitation of the  ionized gas}.  Many  early-type galaxies have optical
emission lines  in their   spectrum.  The  character   of these  lines changes
systematically with luminosity.  In  low-luminosity systems,  the  spectrum is
usually that of \HII\ regions. In luminous galaxies  it corresponds to a liner
spectrum (Sadler   1987). This  indicates  that  the  ionization  mechanism is
different in these two types of galaxies.

{\sl Star  formation history}.  The star  formation history  appears to change
systematically with luminosity. For example, the relative abundance of Mg with
respect to Fe correlates with velocity dispersion. Luminous galaxies typically
have [Mg/Fe] $\sim$0.4,  while  fainter galaxies have  values around  0.  This
indicates that the  enrichment history of the  ISM changes systematically with
luminosity.     Lower-luminosity galaxies also show a    larger  spread in the
Mg-$\sigma$ relation  (e.g., Bender 1996), again pointing  to a different star
formation history.    Many low-luminosity  ellipticals   in fact  display star
formation in the    central parts of   the  galaxies.  It  appears that  disky
galaxies have stronger H$\beta$ indices,  indicating that some star  formation
occurred recently (de Jong \& Davies 1997).

{\sl X-ray emission}.   The amount of  X-ray emission correlates strongly with
optical  luminosity.   In  large   ellipticals, part   of the  X-ray  emission
originates from a halo of hot  gas, while in  smaller ellipticals the X-ray is
due only to X-ray binaries (e.g., Canizares et al.\ 1987).

Many   of these differences  between  different galaxies can   be explained by
different amounts of  gas  present in the   formation/evolution, and  in  many
models for galaxy  formation the gas supply is  a key factor  (e.g., Kauffmann
1996). For example,   the differences between  boxy  and  disky galaxies,  the
importance  of rotation vs.\ anisotropic  galaxies,  and the different central
density distributions are possible a consequence of the relative importance of
gas.  Obviously,  since stars  form   from gas, the  different star  formation
histories must be related to different gas contents during the evolution.

Considering the relation  between gas and  these  different properties,  it is
worthwhile to investigate the systematics of the \HI\ properties of early-type
galaxies.

\section{\HI\ content}

Before discussing the \HI\ properties of early-type  galaxies, it is important
to define which type  of galaxies are  considered and in which environment the
galaxies are.  Table   1 lists the  detection  rates  for different   kinds of
early-type galaxies. A few things  are evident from  this table. First, only a
small fraction of  `pure' elliptical galaxies  have detectable amounts of \HI.
But as soon as the optical morphology shows  some peculiarity, the probability
of detecting    \HI\ increases  dramatically.   This   result  has often  been
interpreted  to mean that  the  origin of  the  \HI\ in elliptical galaxies is
external (e.g.\  Knapp et al.\  1985). It implies that   if we investigate the
characteristics of the \HI\ in these galaxies (morphology and kinematics), one
is considering a subset of the whole population of early-type galaxies, namely
those for which it  is likely  that  some interaction/accretion in  the recent
past has occurred, and it  is important to keep  this in mind. Nevertheless it
is important  not to restrict samples  to  `pure' ellipticals with  no optical
peculiarities, since the \HI-rich galaxies may represent an important phase in
the evolution of many early-type galaxies.

The table does however suggest that there may be a  second origin for the \HI\
in early-type galaxies.  The detection rate also depends  strongly on how much
stellar  disk   is present in a  galaxy.   This could imply    that not in all
early-type   galaxies   the  presence  of \HI\  is     due to a   {\sl recent}
accretion.  The  fact that  the  \HI\ content is   related  to the fundamental
structure of a galaxy could suggest that some of the disky galaxies may have a
long-lived ISM. It appears that many  early-type galaxies, especially those in
the field, indeed often have an ISM with similar characteristics as the ISM in
spirals, the  main difference being that  early-type galaxies have less  of it
(see e.g., Knapp 1998).

\begin{figure}
\centerline{\psfig{file=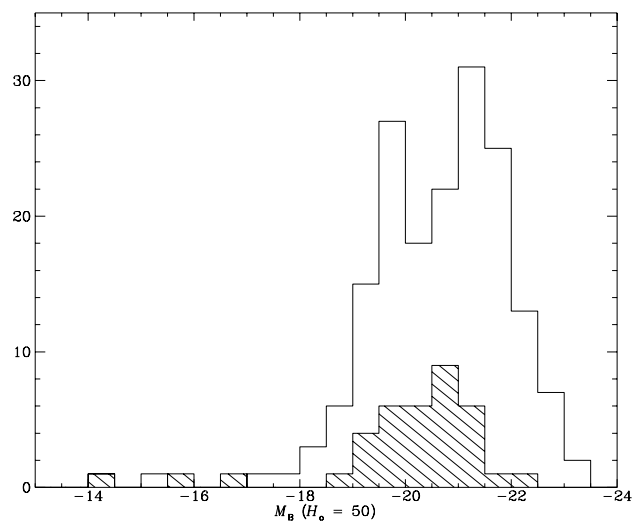,width=10cm}}
\caption{\HI\ detection rate of E and E/S0 galaxies as function of absolute
magnitude. The shaded histogram gives the distribution  of the detections, the
unshaded histogram of the non-detections.  Data from Bregman et al.\ 1992}
\end{figure}

It is often stated that low-luminosity early-type  galaxies are more likely to
have \HI.  This is usually based  on a study done  by Lake and Schommer (1984)
who  observed a small sample of  low-luminosity early-type  galaxies and found
that the detection rate of their sample was significantly  higher than that of
more luminous early-type galaxies as it was known at  the time.  However, when
using larger samples  of \HI\ data on  early-type galaxies that  are available
now,  the situation appears  to be  somewhat different from  that suggested by
Lake and Schommer.

\begin{figure}
\centerline{\psfig{file=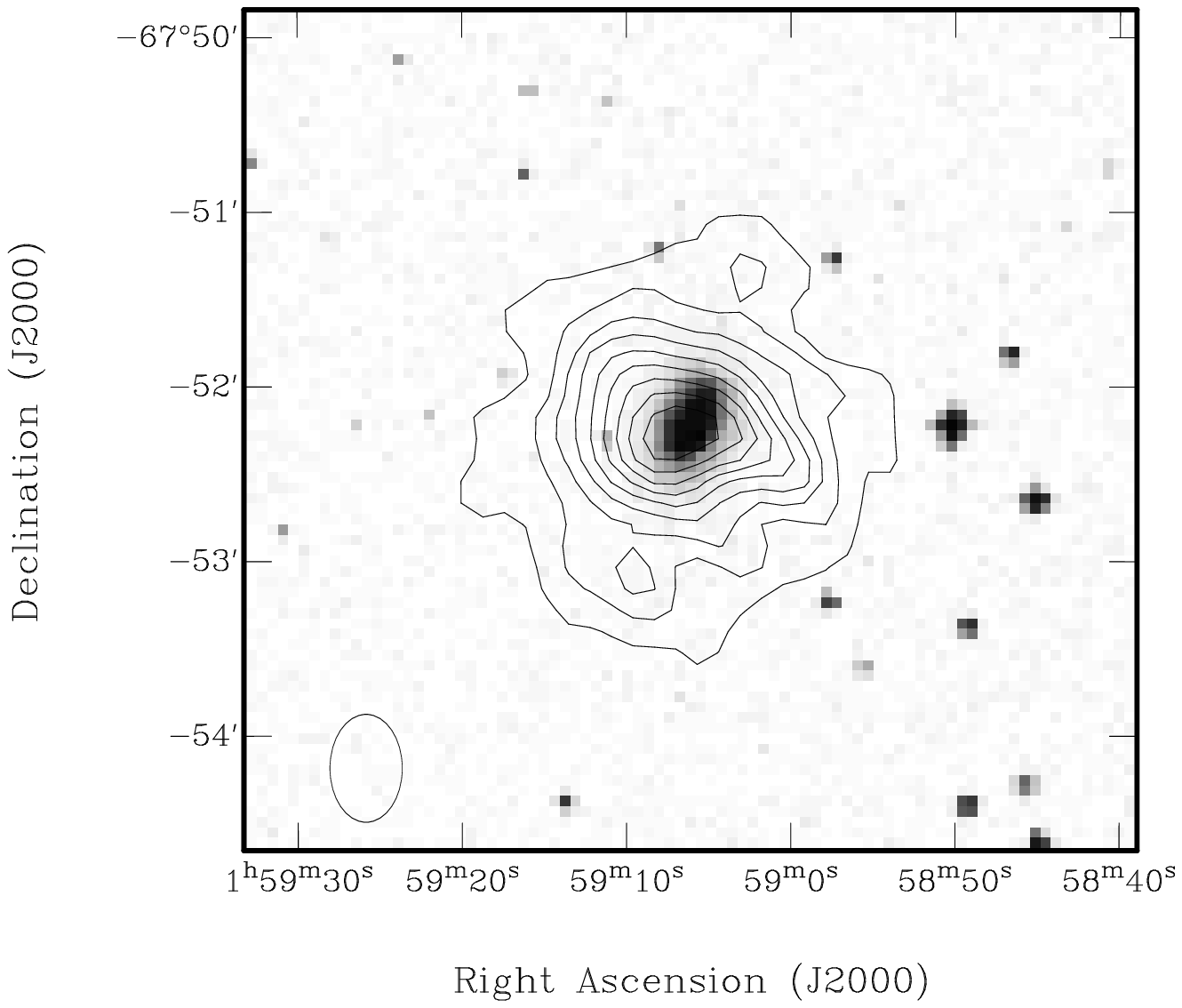,height=8cm}
\hss
\psfig{file=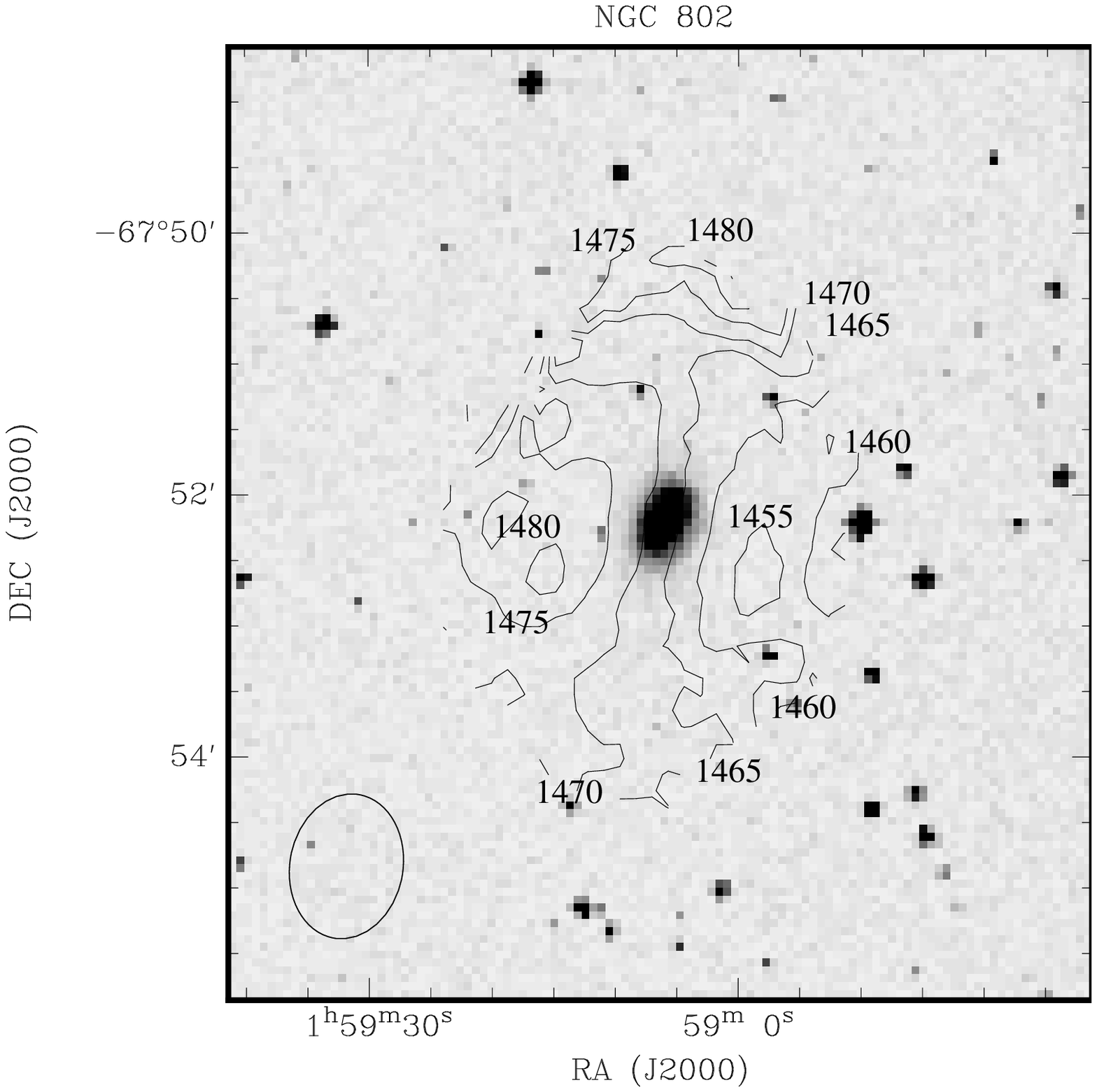,height=8cm}}
\caption{{\sl left:} Contours of the 
total \HI\ image of NGC 802 as obtained with ATCA using
uniform weighting. 
{\sl right:} \HI\ velocity field of NGC 802 from the natural weighted ATCA
data. Contour values (heliocentric, \kms) are indicated. 
The optical image in both figures is taken  from the Digital Sky Survey}
\end{figure}

In figure 1 we show the detection rates of early-type  galaxies as function of
luminosity as it can be  derived from the  compilation of  data of Bregman  et
al.\ (1992).  The figure shows that  galaxies brighter than absolute magnitude
$-22$ appear to  be  poorer in \HI\  than  galaxies fainter than  this  limit.
However, for galaxies  in the magnitude  range $-16$ to $-21.5$, the detection
rate appears   to  be reasonably   flat.  There is   no  strong evidence  that
low-luminosity galaxies (absolute   magnitude  between $-16$ and  $-19$)   are
richer in \HI\ than    galaxies in the range   $-19$   to $-22$.  A    similar
conclusion was  obtained  by  Knapp et al.\   (1985). It  appears that  a more
correct statement about \HI\ content would  be that the most luminous galaxies
are poor in \HI.  Galaxies fainter than $M_B =  -16$ appear to have more often
\HI, although the number  of galaxies for which  data are available is  small.
It is   however,  somewhat difficult to  derive  strong  conclusions from  the
compilation of  Bregman  et al., because it  consists  of a  mix of  field and
cluster  galaxies, and   differential  environmental  effects   are   possibly
important.

\section{\HI\ Morphology and Kinematics}

One interesting difference between low-luminosity galaxies ($-19 > M_B > -16$)
compared to more  luminous galaxies ($-22 >  M_B > -19$) is  that the range of
morphology and  kinematics observed is different for  the two groups. From our
data, together with data in the literature (e.g.\ Lake et al.\ 1987) there are
now   about  10 \HI\ data   cubes    available for low-luminosity   early-type
galaxies. Almost without exception,  the \HI\ in these galaxies  is in a  disk
with a  regular morphology and kinematics.  In some galaxies there is evidence
that (part of)  the  \HI\ may have  been  accreted  recently, but  in  several
galaxies the structure of  the \HI\ is  very regular and  there is no evidence
from the kinematics that a recent accretion has  occurred. To illustrate this,
in figure 2 we give the total \HI\ image and the velocity field for the galaxy
NGC 802  ($M_B   =  -18$).  The   \HI\  in  this  galaxy   is  very  centrally
concentrated. The velocity field shows however that the gas is rotating around
the optical major axis,  like in polar-ring  galaxies. This suggests  that the
\HI\ in NGC 802    has  been accreted   after  the   main stellar body     had
formed. Another galaxy  low-luminosity galaxy that  shows a very  similar \HI\
configuration is NGC 855 (Walsh et al.\ 1990).

\begin{figure}
\centerline{\psfig{file=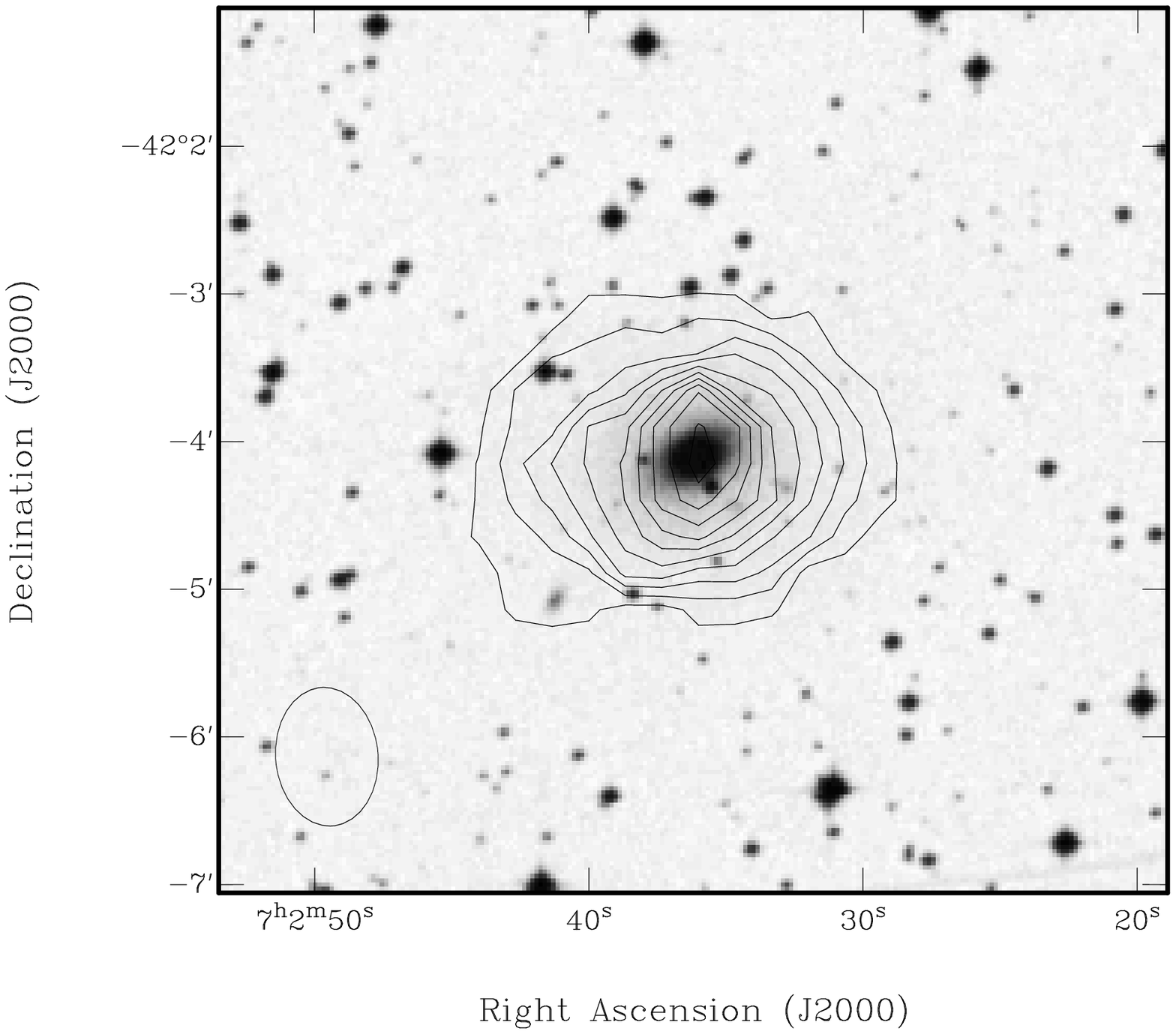,width=9cm}
\hss
\psfig{file=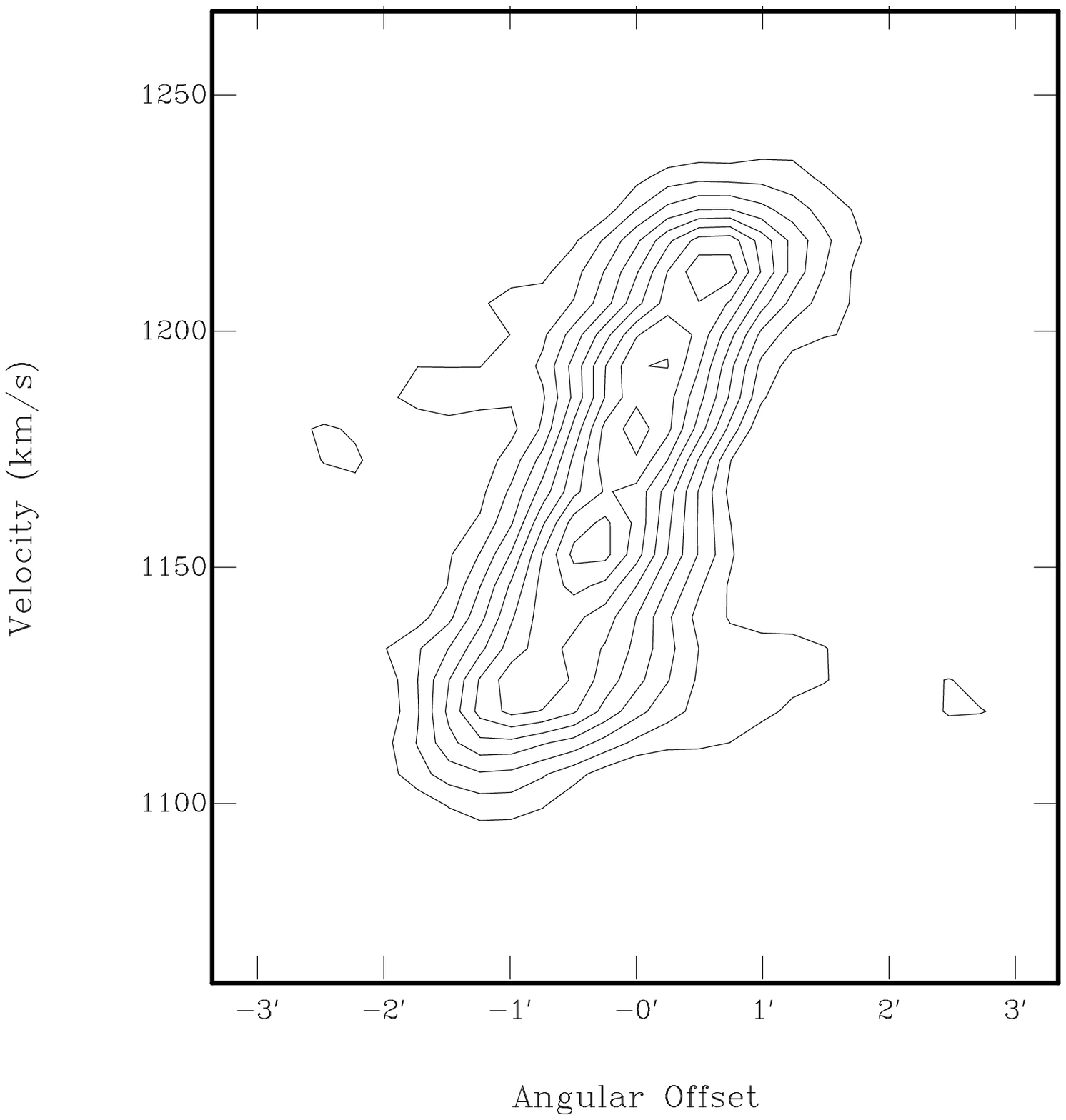,width=8cm}}
\caption{{\sl left:} Total \HI\ image of NGC 2328 as obtained with ATCA. The 
optical image  is taken  from the Digital Sky Survey.
{\sl right:} Position-velocity map taken along the major axis of NGC 2328}
\end{figure}

An example of a   regular \HI\ disk in a   low-luminosity galaxy is given   in
figure 3, where we show the total \HI\ image and a position-velocity map taken
along  the major axis  of this galaxy. Also here,  the \HI\ is quite centrally
concentrated. The position-velocity map in figure 3 shows that the \HI\ in NGC
2328 is in a regularly rotating disk, aligned with the optical body.

In contrast  to the low-luminosity galaxies,  the range in \HI\  morphology in
the more  luminous galaxies ($-22  > M_B  >  -19$) is  much broader.  For this
group,  in most galaxies  the \HI\ shows   an irregular morphology, indicating
that the  gas  is accreting onto the   galaxy, or is  left over  from a recent
merger event. A good example of this  is NGC 5266  (figure 4; Morganti et al.\
1997). This is  a minor-axis dust-lane elliptical with  a large amount of \HI\
($\sim$$10^{10}$ \Msol, $M_{\rm HI}/L_B \sim 0.2$).  Almost all the \HI\ is in
an elongated structure parallel to the optical major axis. Most of this gas is
rotating in a reasonably  regular fashion, although  several subsystems can be
identified    that are not in   stable  circular rotation. Interestingly, this
large-scale \HI\  structure is  perpendicular to  the  inner, minor axis, dust
lane,  and some \HI\  associated with this  dust lane is  also detected (a few
percent of the \HI\ mass). Clearly, NGC 5266 is  a system where a large amount
of  \HI\ has been accreted  recently, or is a  remnant of a recent merger, and
the \HI\ is still settling in the galaxy.

Interestingly, there are now a few luminous  early-type galaxies known that do
have {\sl very regular} \HI\ structures. A very  good example is the E4 galaxy
NGC 807  (figure  5). Deep  \HI\ observations  reveal a low-surface brightness
\HI\ disk    that  shows  no  signs   that   this  \HI\  has   been   accreted
recently. Figure 5 gives the position-velocity  map of this \HI\ disk, clearly
showing the regular rotation of this disk. The evolution of  this disk is very
slow, and this \HI\ disk    can be quite   old.   Often, these regular    \HI\
structures have  a depression or hole  in the centre that  is filled up with a
disk of ionized gas that has very similar kinematics as the  \HI\ disk. A good
example of this is the dust-lane galaxy NGC 3108.

\begin{figure}
\centerline{\psfig{file=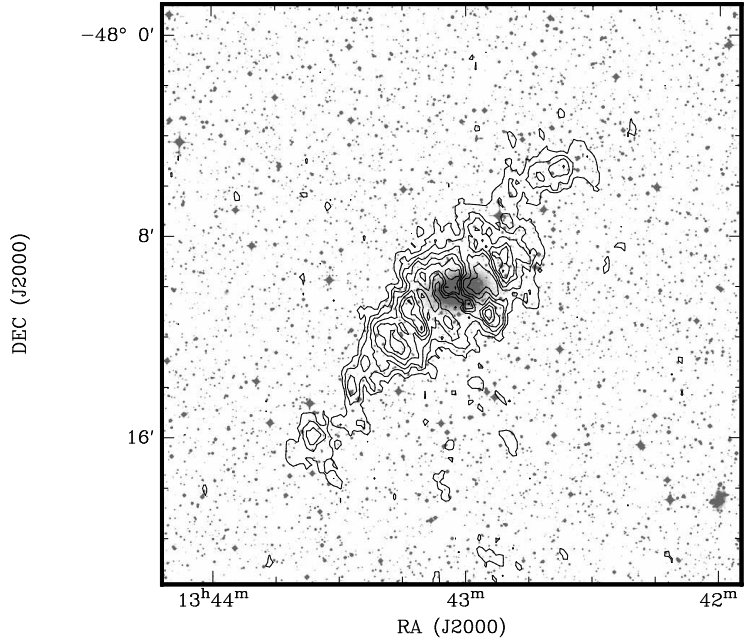,width=10.5cm,angle=90}}
\caption{ Contours of the total \HI\ image of NGC 5266 superposed on an
optical image obtained from the Digital Sky Survey}
\end{figure}

\begin{figure}
\centerline{\psfig{file=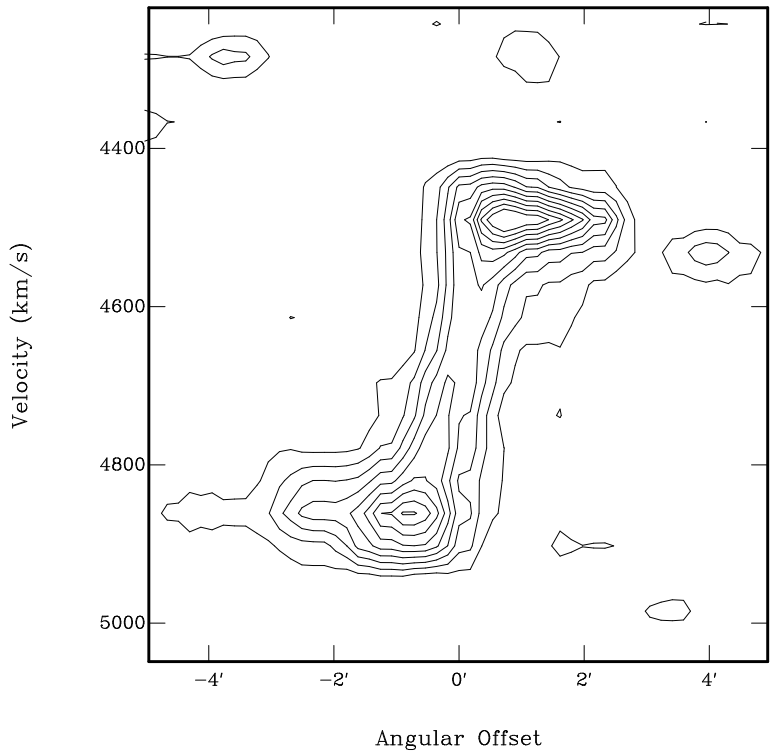,width=10.5cm}}
\caption{ Position-velocity plot of NGC 807,
taken along the kinematical major axis (data taken from the VLA archive,
original data taken by Dressel in 1985)}
\end{figure}

Another striking difference between the \HI\ structures seen in low-luminosity
early-type galaxies and in more luminous galaxies, is that the central surface
brightnesses are quite different. In the low-luminosity  galaxies, the \HI\ is
quite centrally concentrated with central \HI\ surface densities of at least 4
\Msol\,pc$^{-2}$.  These densities are high enough for star formation to occur
on  a reasonable large scale,   and indeed star   formation is observed in the
centres of these  galaxies. Outside the  centre, the surface  densities of the
\HI\ are below  1  \Msol\,pc$^{-2}$ and  perhaps only sporadic  star formation
could occur. In contrast, the surface  densities in the more luminous galaxies
are much  lower, even in the  galaxies with a regular  \HI\ disk or ring.  The
peak surface  densities are typically around 1  \Msol\,pc$^{-2}$,  too low for
large scale star formation to occur.  Figure 6  shows the \HI\ density profile
of  a  low-luminosity  elliptical  and of  a    more luminous E4   galaxy. The
difference between these  profiles is  quite  typical for what  is observed in
most galaxies.

\section{Connection with other properties}

The  range of \HI\ properties observed  in early-type galaxies is quite large,
but  it appears that there are  a  few systematic trends in  the  data, and in
particular  that some of  the \HI\ properties may   be connected to properties
observed at other wavebands.

Many low-luminosity early-type  galaxies that  have \HI, have  this  \HI\ in a
regularly rotating disk.  In the  optical,  these galaxies also show  a  disky
morphology and are  rotationally supported. One possibility  is that the  \HI\
disk observed is the normal  gas counterpart of the  stellar disk structure in
these galaxies. The central surface densities of  the \HI\ are high enough for
star formation to occur, and indeed star  formation is observed in the centres
of many of these  galaxies, and the optical  spectrum of the emission lines is
that of \HII\ regions. The higher densities of the \HI\  could also be related
to the  steeper cores that  are observed in  low-luminosity galaxies, although
also other mechanisms could be responsible for that.  It appears that the \HI\
properties of low-luminosity early-type  galaxies fit in with other properties
of these galaxies, indicating  that the  ISM  in these galaxies has  played an
important role in determining the structure of these galaxies.
\begin{figure}
\centerline{\psfig{file=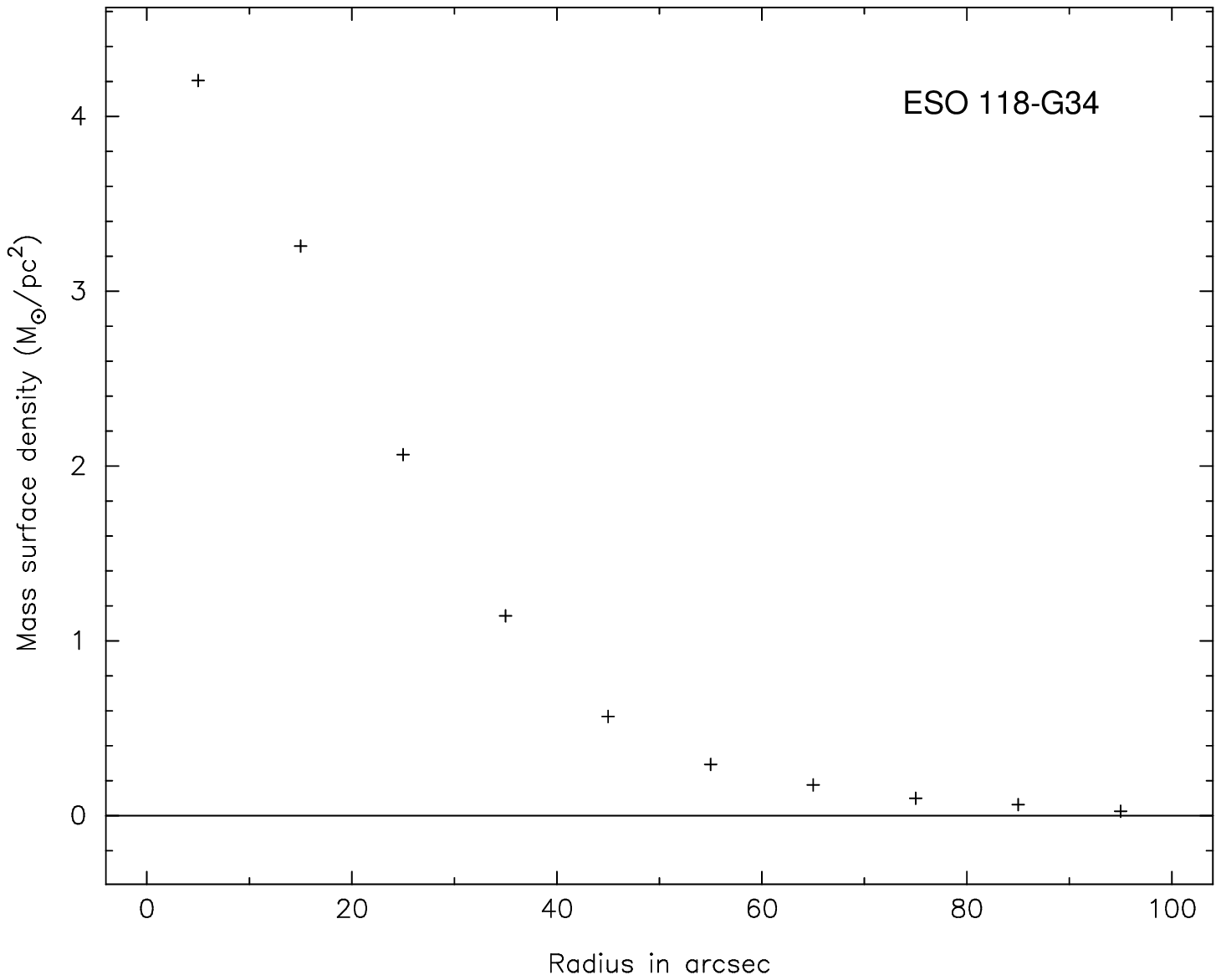,width=9.5cm,angle=-90}
\hss
\psfig{file=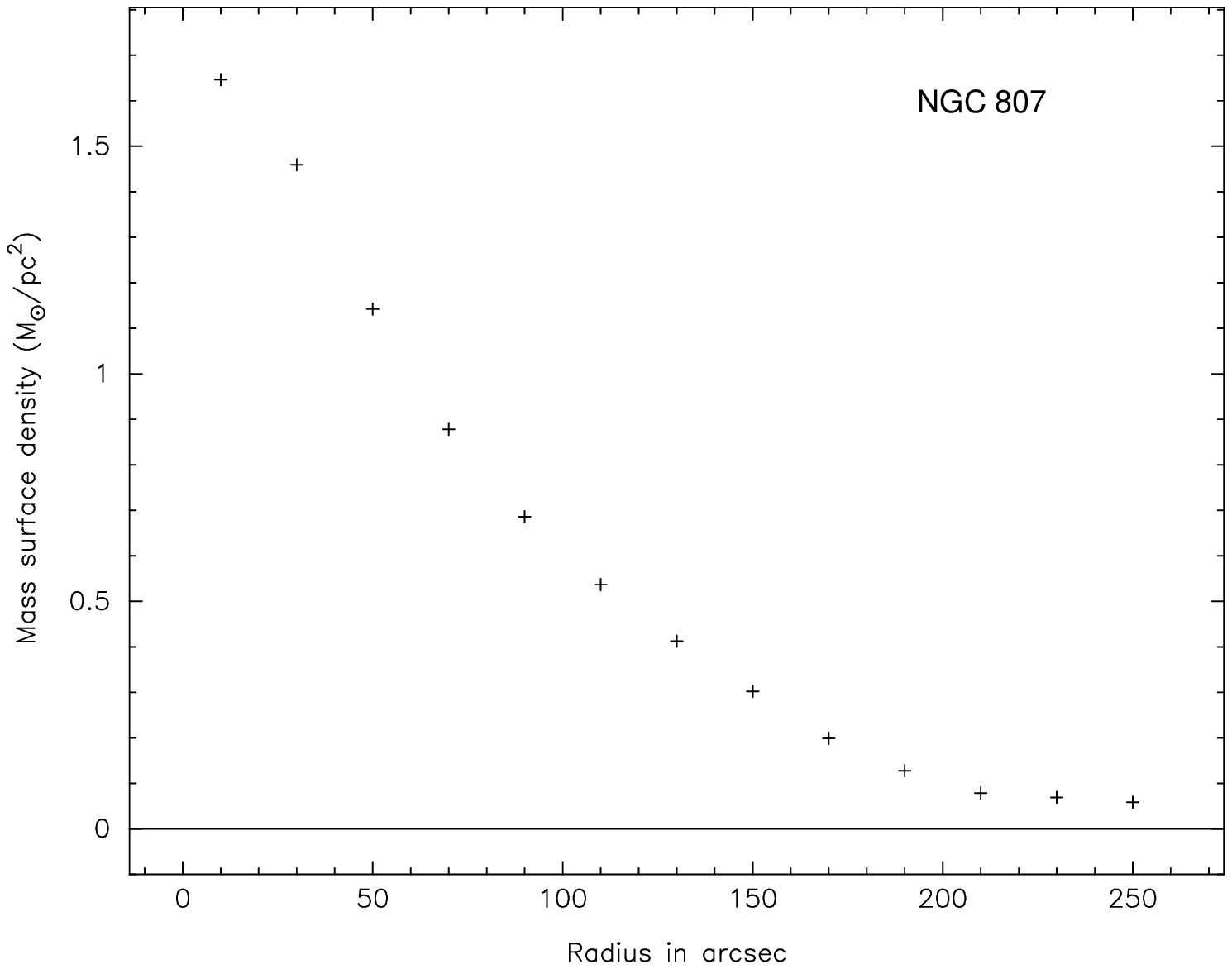,width=9.5cm,angle=-90}}
\caption{Radial \HI\ surface density profiles of the low-luminosity elliptical
ESO 118--G34 {\sl (left)} and the luminous E4 galaxy NGC 807 {\sl (right)}}
\end{figure}
The regular \HI\  disks/rings observed in  the more luminous galaxies could be
similar in origin and  character to the ones observed  in the smaller galaxies
(e.g.\ Morganti et al.\ 1998b), except that some mechanism must be responsible
for keeping the  central  surface density of  the   \HI\ low. A clue   to this
mechanism could be  that  the centres  of  these regular structures  are often
filled up  with  a disk  of ionized  gas  that shows similar   kinematics. The
conditions appear to be such that high \HI\ surface  densities cannot build up
because the \HI\ in  the centre gets  ionized. This could  be connected to the
fact that more luminous early-type galaxies often have a halo of hot has, that
could interact with the \HI\ and ionize it (e.g.\ Goudfrooij 1998). This could
also explain the different excitation of the optical gas that is observed.

In a few low-luminosity  galaxies, there is still evidence  that the  \HI\ has
accreted recently, a process that appears to occur more often in more luminous
galaxies. The different characteristics of the \HI\  in these galaxies suggest
that the  accretion of the \HI\  occurs in a  different  way in low-luminosity
galaxies   compared to the more   luminous ones.   In low-luminosity galaxies,
acccretion appears to result in a more regular \HI\ structure. This could also
be due  to interactions with  a  halo of  hot gas playing   a role in luminous
galaxies. If  \HI\ falls into  a more  luminous   galaxy, the \HI\  could  get
partially ionized and may not have time to settle in a disk-like structure. In
NGC 4696 such an interaction could be occurring (Sparks  et al.\ 1989; de Jong
et  al.\   1990), although this  galaxy is   in a cluster  and   it may not be
representative for the galaxies we have studied.

Another  factor   affecting   the   way  gas  is     accreted   could  be  the
environment. Several of  the more  luminous  galaxies we studied  are in small
groups  of    galaxies,   while    the low-luminosity   galaxies    are   more
isolated.   Interactions and accretions  are of  course  more  common in small
groups and  less relaxed  \HI\  structures should be    more common. Also  the
luminous  galaxies  with regular  \HI\  structures tend   to be more isolated,
consistent  with  the idea that environment  plays  an  important role  in the
evolution of \HI\ in early-type galaxies.

\bigskip
{\sl Acknowledgements.}  The optical images shown in this paper are taken from
the Digital  Sky Survey. These image are  based on photographic  data obtained
using The UK  Schmidt Telescope. The UK  Schmidt Telescope was operated by the
Royal Observatory Edinburgh, with funding from  the UK Science and Engineering
Research  Council, until  1988 June, and   thereafter by the  Anglo-Australian
Observatory.  Original plate material  is  copyright (c) the Royal Observatory
Edinburgh and the Anglo-Australian Observatory. The plates were processed into
the  present compressed digital form with  their permission. The Digitized Sky
Survey  was  produced  at  the  Space  Telescope Science   Institute under  US
Government grant NAG W-2166.

\section*{References}

\reference Bender, R. 1996, in New Light on Galaxy Evolution, IAU symp. 171,
eds R. Bender, R. Davies, Kluwer, p181

\reference Bregman, J.N., Hogg, D.E., Roberts, M.S.  1992, ApJ, 387, 484

\reference Canizares, C.R., Fabbiano, G., Trinchieri, G. 1987, ApJ, 312, 503

\reference Davies R.L., Efstathiou G., Fall S.M., Illingworth G.D., Schechter
P. 1982, ApJ, 266, 41

\reference de Jong, T., Norgaard-Nielsen, H.U., Jorgensen, H.E., Hansen, L.  
1990, A\&A, 232, 317

\reference de Jong, R., Davies, R. 1997, MNRAS, 285, 1

\reference Goudfrooij, P.  1998, in Star Formation in Early-Type Galaxies,
eds. P. Carral and J. Cepa, ASP Conf. Proc., in press (astro-ph/980957)

\reference Kauffmann, G. 1996, MNRAS, 281, 487

\reference Knapp, G.R., Turner, E.L., Cunniffe, P.E. 1985, AJ, 90, 54

\reference Knapp, G., 1998, in Star Formation in Early-Type Galaxies,
eds. P. Carral and J. Cepa, ASP Conf. Proc., in press (astro-ph/9808266)

\reference Lake, G., Schommer, R.A. 1984, ApJ, 280, 107

\reference Lake, G., Schommer, R.A., van Gorkom, J.H. 1987, ApJ, 314, 57

\reference Lauer, T. 1997, in Second Stromlo Symposium: The Nature of Elliptical
      Galaxies, eds M. Arnaboldi, G.S. Da Costa P. Saha, eds, ASP Conf. Ser. Vol
      116, p.113

\reference  Morganti, R., Sadler, E., Oosterloo, T., Pizzella, A., 
Bertola, F., 1997a, AJ, 113, 937

\reference Morganti R., Sadler E., Oosterloo T., 1997b, in: Second Stromlo
Symposium: The Nature of Elliptical Galaxies, M.  Arnaboldi, G.S.  Da Costa,
P. Saha, eds, ASP Conf. Ser., Vol 116, p. 354

\reference Morganti R., Oosterloo T., Tsvetanov Z., 1998a, AJ, 115, 915

\reference Morganti R., Oosterloo T., Sadler E.M., Vergani D., 1998b, in: Star
formation in early-type galaxies, ASP Conf. Ser., in press

\reference Sadler, E.M 1987, in: Structure and Dynamics of Elliptical
Galaxies, IAU symp. 127, p125 

\reference Sparks, W.B., Macchetto, F.,  Golombek, D. 1989, ApJ, 345, 153

\reference  Walsh, D.E.P., Van Gorkom, J.H., Bies, W.E., Katz, N., Knapp, G.R.,
 Wallington, S. 1990, ApJ, 352, 532

\end{document}